# Microwatts continuous-wave pumped second harmonic generation in few- and mono-layer GaSe

Xuetao Gan[1,*], Chenyang Zhao,[1] Siqi Hu,[1] Tao Wang,[2] Yu Song,[1] Jie Li,[2] Qinghua Zhao,[2] Wanqi Jie,[2] and Jianlin Zhao[1,*]

[1]*MOE Key Laboratory of Space Applied Physics and Chemistry, and Shaanxi Key Laboratory of Optical Information Technology, School of Science, Northwestern Polytechnical University, Xi'an 710072, China*
[2]*State Key Laboratory of Solidification Processing, Northwestern Polytechnical University, Xi'an 710072, China*
E-mail: xuetaogan@nwpu.edu.cn; jlzhao@nwpu.edu.cn

**Abstract**

We demonstrate the first achievement of continuous-wave (CW) pumped second harmonic generation (SHG) in few- and mono-layer gallium selenide (GaSe) flakes, which are coated on silicon photonic crystal (PC) cavities. Because of ultrahigh second order nonlinearity of the two-dimensional (2D) GaSe and localized resonant mode in the PC cavity, SHG's pump power is greatly reduced to microwatts. In a nine-layer GaSe coated PC cavity, while the optical power inside the GaSe flake is only 1.5% of that in the silicon PC slab, the SHG in GaSe is more than 650 times stronger than the third harmonic generation in silicon slab, indicating 2D GaSe's great potentials to strengthen nonlinear processes in silicon photonics. Our study opens up a new view to expand 2D materials' optoelectronic applications in nonlinear regime and chip-integrated active devices.

## INTRODUCTION

Two-dimensional (2D) materials with layered structure, such as graphene, black phosphorus, transition metal dichalcogenides and gallium monochalcogenides, have attracted much recent attentions to complement silicon electronics and optoelectronics potentially [1-3]. Second harmonic generation (SHG), normally achieved in noncentrosymmetric materials, as one of the strongest nonlinear processes has been widely employed in signal and image processing, laser industries, and optical spectroscopies [4, 5]. In virtue of shapeable electronic structures in the atomic layered crystals, SHGs in 2D materials occupy a variety of distinct characteristics. For instance, while the bulk $MoS_2$ and *h*-BN have centrosymmetry, their finite slices with odd-layer thicknesses could yield moderately strong SHG, especially for the mono-layers with pure noncentrosymmetric $D_{3h}$ space group [6-8]. In contrast, SHGs were observed in even-layered 1T' $MoTe_2$ determined by the broken inversion symmetry [9]. SHGs in 2D materials also indicate unusually large second order nonlinear susceptibilities, which are nearly three orders of magnitude higher than other common nonlinear crystals [10, 11]. At 2D crystals' boundaries, variations of electronic structure result in even stronger SHG with one-dimensional edge state, facilitating direct optical imagings of crystal orientations and straightforward reveals of kinetic nature of grain boundary formation in as-grown mono-layers [12, 13]. In addition, the strong exciton charging effect in mono-layer semiconductor field-effect transistor allows for exceptional control over the oscillator strengths between the exciton and trion resonances, enabling the electrically controlled nonlinear susceptibilities by an order of magnitude and counter-circular resolved SHG spectrum [14].

During the past few years, 2D materials have been widely reported for promising optoelectronic devices including photodetectors, modulators, and light emission devices [15-21]. If 2D materials' extraordinary SHGs were exploited further, their optoelectronic applications might be greatly extended into nonlinear regimes for coherent light source generations, image processings, ultrafast laser engineerings, etc. However, to the best of our knowledge, the previously demonstrated SHGs in 2D materials all required pulsed laser sources with high peak power (larger than 100 mW). For practical applications, continuous-wave (CW) pumped SHG by means of simple, low-power and low-cost light sources, such as semiconductor laser diodes, would be highly desirable.

In this letter, we report it is possible to achieve strong SHG in atomically layered GaSe using a low-power CW laser. GaSe is a well-known nonlinear crystal in the spectral range from visible to terahertz [22]. It has three most important classifications according to the layer stacking order, i.e., β-GaSe, γ-GaSe, and ε-GaSe. Here, we employ ε-GaSe, whose stacking sequence leads to the absence of inversion center for arbitrary layer thickness [23]. Hence, the achievement of SHG in GaSe has no layer dependence. In addition, GaSe has pronounced second order nonlinearity in its primitive layer consisting of two atomic layers of Ga sandwiched between two atomic layers of Se, which is 1-2 orders of magnitude larger than that of mono-layer $MoS_2$ [11]. By integrating few- and mono-layer GaSe flakes with silicon photonic crystal (PC) cavities to enable an effective interaction between GaSe and cavity's evanescent field, a telecom-band CW laser is strong enough to pump SHG in the 2D GaSe, and the required power could be reduced down to only few microwatts. In addition, we observe the cavity-enhanced SHG in a nine-layer GaSe is more than two orders of magnitude stronger than the cavity-enhanced third harmonic generation (THG) in a thick silicon slab. It indicates GaSe's high second order nonlinearity could enable significant chip-integrated nonlinear effects. Therefore, integration of 2D GaSe, with high nonlinearity and negligible linear and nonlinear losses in telecom-band, onto nanophotonic structures may pave a new way towards developing high-performance chip-integrated nonlinear devices based on second order nonlinearity, such as on-chip frequency-conversion laser sources [24], optical autocorrelator [25], entanglement photon-pair generations [26].

**MATERIALS AND METHODS**
We fabricated PC cavities on a silicon-on-insulator (SOI) wafer with a 220 nm thick top silicon layer (see Supporting Information), whose fabrication process is CMOS (Complementary Metal Oxide Semiconductor)-compatible and the integrated devices have potentials for future on-chip optical interconnects. Also, silicon has no second order nonlinearity, ensuring the measured second harmonic signal is purely from the GaSe layer without a background from the bulk silicon substrate. The exfoliated GaSe flake is integrated onto the PC cavity using a dry transfer method (see Supporting Information) [27]. Figure 1(a) displays an optical microscope image of the fabricated device with a large few-layer GaSe flake covering the cavity. The uniformity and thickness of the GaSe flake are examined using an atomic force microscopy (AFM), as shown in Fig. 1(b). The profile heights over the PC air-holes indicate conformal contract between the few-layer flake and the PC cavity with the dry transfer technique, which promises the effective interaction between GaSe and cavity evanescent field. The thickness of the transferred GaSe is examined as 7.8 nm, corresponding to a layer number of nine by assuming the mono-layer thickness of 0.85 nm [28]. The photoluminescence (PL) of the transferred nine-layer GaSe is evaluated with a pump of 532 nm CW laser, as shown in the inset of Fig. 1(a). Because of the direct-to-indirect and broadening bandgap transition in the few-layer GaSe, the PL peak locates at 587 nm with a greatly reduced quantum yield. To achieve small mode volume and high coupling efficiency of the PC cavity, which is crucial for cavity-enhanced SHG [25, 29-32], we design the cavity defect into a point-shifted type with shrunk air-holes [33], as shown in the scanning electron microscope (SEM) image of Fig. 1(c). Pored holes in the GaSe layer over several PC air-holes are induced by the accelerated electron beam, which is witnessed during the SEM measurements.

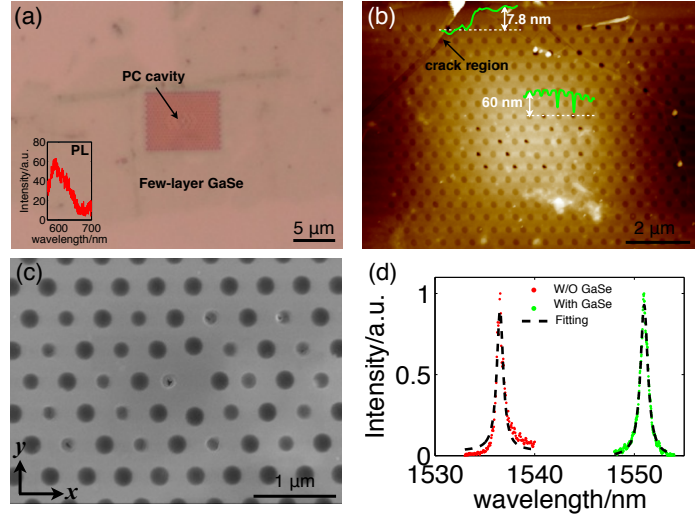

Fig. 1 (a) Optical microscope image of the GaSe-PC cavity with the GaSe layer shown as light green flakes. Inset displays the PL spectrum of the nine-layer GaSe flakes. (b) AFM image of the GaSe-PC cavity, where the thickness of the GaSe layer is indicated at the location of a crack region with a value of 7.8 nm and the dashed lines indicate the locations where the height distributions is measured. (c) SEM image of the GaSe-PC cavity. (d) Cavity's refection spectra of the resonant mode before and after the integration of GaSe.

The experimental measurements over the GaSe-PC cavity are implemented using a vertically coupled cross-polarization microscope (see Supporting Information) [34]. The orthogonal polarizations of the excitation laser and the collection signal are achieved using a polarized beam splitter (PBS). A half wave plate (HWP) is inserted between the PBS and the objective lens to control the laser polarization with respect to the axis of the PC cavity. To evaluate resonant modes of the PC cavity, we illuminate the cavity with a CW tunable narrowband laser and monitor its reflection using a telecom-band photodiode. Cavity reflection spectra could be obtained by scanning the laser wavelength with a step of 0.005 nm.

**RESULTS AND DISCUSSIONS**

The red and green dotted lines in Fig. 1(d) display the same resonant mode of the PC cavity measured before and after the integration of GaSe, and the black dashed lines are Lorentizan fittings. After the integration, the GaSe layer works as a positive perturbation of the dielectric function around the resonant mode, which shifts the resonant wavelength from 1536.5 nm to 1551.0 nm. It also weakens the confinement of the resonant mode, lowering the $Q$ factor from $\sim$2,000 to $\sim$1,750. These variations of the resonant mode are confirmed using a three-dimensional finite element simulations (COMSOL Multiphysics) with the refractive index and thickness of the GaSe layer chosen as 2.8 and 7.8 nm [35].

To implement the cavity-enhanced SHG measurement, we tune the laser wavelength as 1551 nm to excite the cavity's resonant mode, and the reflected resonant signal is filtered out using a short pass dichroic mirror (see Supporting Information). The frequency conversion signal scattered from the cavity is monitored using a 0.5 m spectrometer mounted with a cooled silicon camera. Figure 2(a) displays an obtained spectrum when the pump power is 0.5 mW measured after the objective lens. At the wavelength of 775.5 nm, a strong peak is observed, corresponding to the second harmonic signal of the pump laser. Another weak peak arises as well at 517 nm, equaling to the THG wavelength of the pump laser.

Before we transfer the GaSe layer, we examine the harmonic generations from the bare silicon PC cavity as well, which presents the similar THG peak intensity but no SHG peak. As reported in Ref. [31], resonantly enhanced SHG in a bare silicon PC cavity is also possible due to the surface second order nonlinear process. However, this SHG signal is about two orders of magnitude weaker than the cavity-enhanced THG [31], which is a possible reason for the failed observation of SHG in our bare PC cavity. Hence, the obtained strong SHG from the GaSe-PC cavity is due to the top GaSe layer. Also, the THGs measured before and after integration of GaSe have no noticeable power variation, indicating the THG in Fig. 2(a) mainly results from the third order nonlinear process in the bulk silicon slab. Comparing SHG and THG of the GaSe-PC cavity, the SHG peak is more than 650 times stronger than the THG peak. From the resonant mode simulation, we calculate the optical power distributions in the GaSe-PC cavity, indicating the power inside the nine-layer GaSe flake is only 1.5% of that in the bulk silicon slab. The remarkably strong SHG signal is in consistent with the ultrahigh second order nonlinear susceptibility of GaSe [11, 22, 28].

We also evaluate the pump power dependence of the SHG by varying the laser power, as shown in the log-log plot of Fig. 2(b). The dashed line indicates a fitting with a slope of 2.01, verifying the quadratic power dependence of the cavity-enhanced SHG. By measuring the cavity reflection of the on-resonance laser, and combining the far-field collection efficiency of the resonant mode (~40%) [33], the coupling-in efficiency of the pump laser is estimated as 6%. Therefore, even for a 0.1 mW pump illuminates on the PC cavity, the power coupled into the cavity is lower than 10 microwatts, which can still generate a detectable SHG.

For the previously reported SHG in mono-layer or few-layer 2D materials, pulsed lasers are employed to achieve high peak pump power and increase the SHG output signal. In our GaSe-PC cavity, the successful observation of SHG with a CW pump could be attributed to the strong enhancement of the pump power coupled into the cavity. To verify this, we acquire the SHG powers as we tune the pump laser wavelength cross the cavity resonance, as plotted in Fig. 2(c). As the laser wavelength is away from the resonance, the SHG signal is nearly not detectable due to the weak light-matter interaction. For the on-resonance pump, the laser power coupled into the cavity is enhanced by a factor proportional to $Q$, and the enhancement over the second order nonlinear process has a factor proportional to $Q^2$ [25, 29-31]. Hence, the SHG spectrum could be very well described by the squared Lorentzian lineshape, which is used to fit the fundamental resonance shown in Fig. 1(d).

The cavity-enhanced SHG is verified as well using a spatial mapping of the SHG with an on-resonance pump, as shown in Fig. 2(d). The device is mounted on a 2D piezo-actuated stage with a moving step of 100 nm, and the generated second harmonic signal is measured using a photomultiplier tube (PMT). Since the cavity-enhanced SHG is pumped by the evanescent field of the cavity resonant mode, efficient SHG can only be observed when the pump laser couples into the cavity. Therefore, the spatial position of the detected SHG is determined by cavity's coupling-in region. For the modified point-shifted cavity, light could vertically couple into the cavity effectively around the modification region [33], which has a dimension about 2.5×2 μm$^2$. In the SHG mapping, a similar area with strong signal is observed. Outside the cavity mode-coupling region, the GaSe layer is only pumped by the vertically illuminated laser, which is too weak to yield observed SHG for a CW pump.

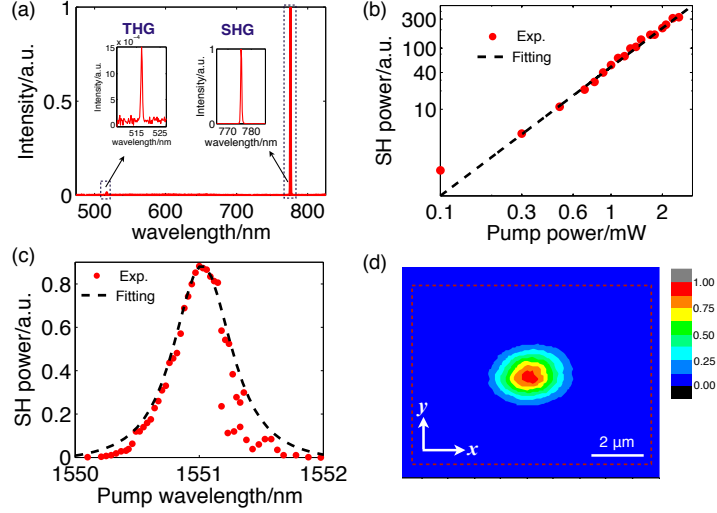

Fig. 2 (a) Spectrum of the cavity's scattering signal with a 1551 nm pump laser, which has two signals at 775.5 nm and 517 nm corresponding to SHG and THG, respectively. (b) Pump power-dependence of the cavity-enhanced SHG with a fitting slope of 2.01, where the pump power is measured after the objective lens. (c) Cavity-enhanced SHG spectrum when the pump wavelength is tuned cross the resonance, where the dashed line is a squared Lorentzian fitting curve of the cavity resonant mode. (d) Spatial mapping of the cavity-enhanced SHG, where the PC boundary is indicated by the dashed line.

Because of the optical losses from various optical components in the microscope setup as well as the absorption by the silicon substrate, it is difficult to evaluate the absolute SHG power enhanced by the cavity. To evaluate the cavity-enhancement factor, we switch the CW laser into a pulsed laser to pump the GaSe-PC cavity, with pumping wavelength (at 1560 nm) being off-resonance from the cavity mode. In this regime, both the on-resonance CW pump and the off-resonance pulsed pump share the same excitation and collection optical paths, as well as the same location of GaSe layer to maintain the dielectric environment of the SHG emission. For a 2.8 nW SHG measured by the PMT, the required powers for the CW laser and pulsed laser are measured as 2.5 mW and 1.56 mW (averaged power), respectively. For both the CW and pulsed laser excitations, the SHG process strongly depends on the strength of electric field. With parameters of the pulsed laser, we can calculate the effective electrical field used to generate SHG in GaSe. Combining with the experimental results and the calculations, we can estimate the enhancement factor as 612 (see Supporting Information).

In Fig. 3(a), we display the simulated electric fields of the resonant mode located at the GaSe layer, decomposed into the two orthogonal components $E_x$ and $E_y$. Because the odd-symmetry of $E_y$ would generate a far-field pattern splitting into a large oblique angle, the coupling between the cavity mode and a $y$-polarized far-field is very small. For a light with $x$-polarization, its far-field coupling with the resonant mode is high due to the vertically directed far-field pattern of $E_x$. Therefore, when an on-resonance laser focuses on the cavity, only the $x$-polarized component can couple into the cavity effectively. If the laser polarization is changed by rotating the HWP, the power coupled into the cavity is proportional to $\sin^2(2\theta)$, where $\theta$ is the angle between the HWP's fast-axis and the $y$-axis of the PC cavity. And the cavity-enhanced SHG, which is proportional to the square of the coupled power, should follow a function of $\sin^4(2\theta)$. Considering the cross-polarization of the experimental setup, the vertically scattered SHG after the HWP is then projected onto the output polarization direction, and the finally collected SHG has a function of $\sin^6(2\theta)$, as plotted in Fig. 3(b). Here, the four peaks have different maximum values. We attribute it to the imperfection of the HWP at the SHG wavelength (the achromatic wavelength range of the HWP is 1200 nm to 1600 nm).

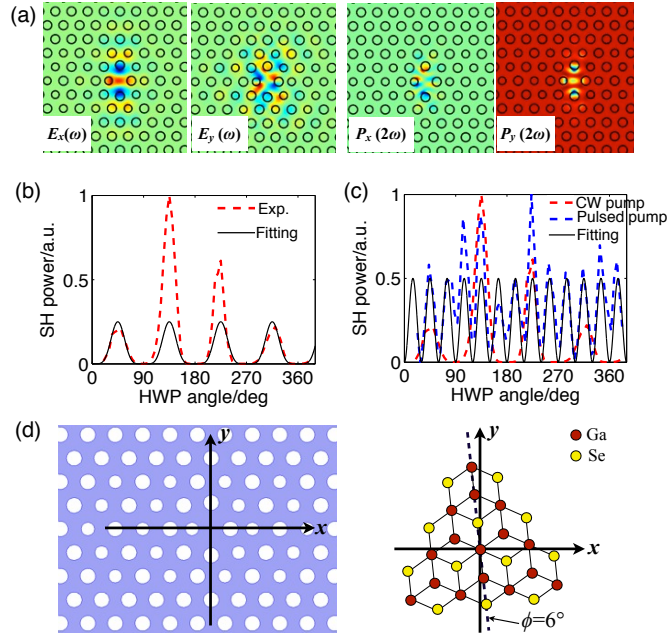

Fig. 3 (a) Simulated electric fields ($E_x$ and $E_y$) of the resonant mode and the corresponding nonlinear polarizations ($P_x$ and $P_y$) calculated from the nonlinear susceptibility matrix and crystal orientation of GaSe. (b) Polarization-dependence of the cavity-enhanced SHG. The black solid line is a fitting of $\sin^6(2\theta)$, where $\theta$ is the angle between the HWP's fast axis and $y$-axis of the PC cavity. (c) Polarization-dependences of the SHG pumped by the on-resonance CW laser and off-resonance pulsed lasers. The SHG pumped by the off-resonance pulsed laser is fitted by a function of $\sin^2 6(\theta+3°)$, indicated by the black solid line. (d) Alignment of the GaSe crystal orientation with respect to the axes of the PC cavity.

By switching the on-resonance CW laser into the off-resonance pulsed laser (at 1560 nm), we observe another pump polarization-dependent SHG from the GaSe-PC cavity by rotating the HWP, as shown in Fig. 3(c). A fitting with a function of $\sin^2 6(\theta+3°)$ is obtained, which is determined by GaSe's $D_{3h}$ symmetry [11]. This 12-fold variation is not observed in the cavity-enhanced SHG, where the GaSe crystal only interacts with the evanescent field of the cavity mode, and the angle between the crystal orientation and cavity's electrical field vector is fixed no matter how the laser polarization change. The polarization variation of the pump laser only changes the laser power coupled into the cavity, which therefore induces the 4-fold variation of SHG following the function of $\sin^6(2\theta)$.

The HWP angle-dependences of GaSe's SHGs pumped by the on-resonance CW laser and off-resonance pulsed laser indicate axes of the PC cavity and GaSe's crystal structure, respectively. We could probe the alignment between GaSe's crystal orientation and axes of the PC cavity by comparing them, as shown in Fig. 3(c). We conclude the Ga-Se bond is aligned to the $y$-axis of the PC cavity with an angle of $\Phi=6°$, as indicated in Fig. 3(d). With this alignment and the near-fields ($E_x$ and $E_y$) of the resonant mode in the GaSe layer, we could calculate the nonlinear polarizations $P_x$ and $P_y$ generated in GaSe according to its nonlinear susceptibility matrix (see Supporting Information), as shown in Fig. 3(a). The symmetries of the generated $P_x$ and $P_y$ indicate the $y$-component of the SHG has a vertical far-field radiation as well as a high coupling efficiency. Considering the employed cross-polarization setup, the $x$-polarized pump laser and $y$-polarized SHG radiation enable the high efficiency generation and collection of SHG.

The low-power CW pumped SHG in 2D materials is also validated in a mono-layer GaSe. Figure

4(a) shows the AFM image of an integrated mono-layer GaSe-PC cavity, which has a resonant mode at the wavelength of 1548.8 nm and a $Q$ factor of ~1,800. By pumping it with a 0.5 mW on-resonance CW laser, we measure the vertically scattered upconversion signal, as displayed in Fig. 4(b). Two clear peaks are observed at wavelengths of 774.4 nm and 516.3 nm, which are verified as GaSe's SHG and silicon's THG, respectively. Figures 4(c) and (d) plot the pump wavelength and polarization dependences of the SHG signal, yielding similar conclusions as those obtained from the nine-layer GaSe-PC cavity. The results also confirm the cavity enhancement effect via resonant mode's near-field. Because the mono-layer GaSe-PC cavity and the nine-layer GaSe-PC cavity have similar $Q$ factors, the cavity-enhanced THGs of the silicon slab from the two devices should be close. Therefore, we could compare the cavity-enhanced SHG powers from the mono- and nine-layer GaSe by calculating the ratios of GaSe's SHG peak to silicon's THG peak of the two devices separately. From the experimental results, the peak ratios from the mono-layer GaSe-PC cavity and nine-layer GaSe-PC cavity are calculated as 8.7 and 650, i.e., SHG from the nine-layer GaSe is about 75 times of that from the mono-layer GaSe. The SHG power variation of the two GaSe flakes is closely consistent with SHG's quadratic dependence on the material thickness for a thin film, as demonstrated in Ref. [28]. While the absolute SHG power of the mono-layer GaSe is much weaker than that of the nine-layer GaSe, the SHG enhancement factors of the two devices are both around 600, which is determined by the $Q$ factor and mode volume of the cavity mode.

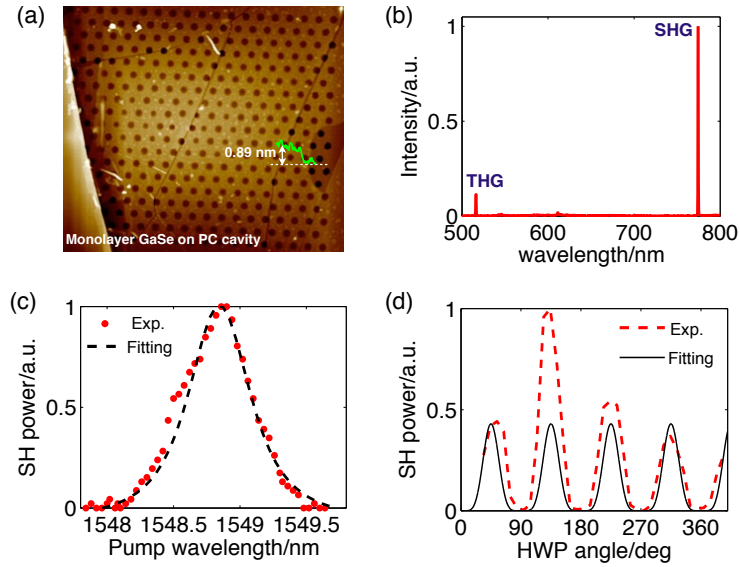

Fig. 4 CW pumped SHG from a mono-layer GaSe integrated with a PC cavity. (a) AFM image of the GaSe-PC cavity, indicating GaSe's thickness of ~0.89 nm. (b) Spectrum of the cavity's upconversion signals pumped with a 1548.8 nm CW laser, showing SHG and THG signals. (c) Cavity-enhanced SHG spectrum when the pump wavelength is tuned cross the resonance, where the dashed line is a squared Lorentzian fitting curve of the cavity resonant mode. (d) Polarization-dependence of the cavity-enhanced SHG.

## CONCLUSIONS

In conclusion, we have presented the first demonstration of CW pumped SHG in 2D materials with the integration of a silicon PC cavity. Assisting by the ultrahigh second order nonlinearity of atomically layered GaSe and extremely confined resonant mode of the cavity, we achieved efficient SHGs in nine- and mono-layer GaSe flakes with an excitation power less than 10 microwatts. Comparing with the SHG pumped by an off-resonance pulsed laser, the on-resonance pumped SHG is enhanced by a factor exceeding 600. The SHG enhancement could be improved further by employing a cavity with even higher $Q$ factor, which is proportional to $Q^2$ [25, 29-31]. On the other hand, a PC cavity with doubly resonant modes at the fundamental pump and second harmonic signal could achieve a significantly high enhancement as well [36]. The CW pumped SHG could open up opportunities for employing 2D materials' distinct second order nonlinearity to extend their optoelectronic devices in nonlinear regime with low optical power, such as coherent laser frequency converters, upconversion detectors, entangled photon pair generators, etc.

## SUPPORTING INFORMATION
Fabrication of the integrated GaSe-PC cavity; Experimental arrangement; Estimation of the SHG enhancement factor by the PC cavity; Nonlinear polarizations in GaSe generated by the cavity mode.

## ACKNOWLEDGEMENT

The authors thank Kaihui Liu for many fruitful discussions. Financial support was provided by NSFC (61522507, 11404264, 61377035, 11634010), and the Natural Science Basic Research Plan in Shaanxi Province of China (2016JQ6004).

## REFERENCES

1. Kim K, Choi J, Kim T, Cho S, Chung H. A role for graphene in silicon-based semiconductor devices. Nature 2011; **479**: 338-344.
2. Li L, Yu Y, Ye G, Ge Q, Ou X, Wu H, Feng D, Chen X, Zhang Y. Black phosphorus field-effect transistors. Nature Nanotech. 2014; **9**: 372-377.
3. Xu M, Liang T, Shi M, Chen H. Graphene-like two-dimensional materials. Chem. Rev. 2012; **113**: 3766-3798.
4. Shen YR. The principles of nonlinear optics. Wiley-Interscience: New York, 2003.
5. Boyd RW. Nonlinear Optics, 3rd ed.; Academy Press: San Diego, CA, 2008.
6. Malard LM, Alencar TV, Barboza APM, Mak KF, Paula AMD. Observation of intense second harmonic generation from $MoS_2$ atomic crystals. Phys. Rev. B 2013; **87**: 201401.
7. Li Y, Rao Y, Mak KF, You Y, Wang S, Dean CR, Heinz TF. Probing symmetry properties of few-layer $MoS_2$ and h-BN by optical second-harmonic generation. Nano Lett. 2013; **13**: 3329-3333.
8. Zhao M, Ye Z, Suzuki R, Ye Y, Zhu H, Xiao J, Wang Y, Iwasa Y, Zhang X. Atomically phase-matched second-harmonic generation in a 2D crystal. Light: Science & Applications. 2016; **5**: e16131.
9. Beams R, Cancado LG, Krylyuk S, Kalish I, Kalanyan B, Singh AK, Choudhary K, Bruma A, Vora PM, Tavazza F, Davydov AV, Stranick SJ. Characterization of few-layer 1T' $MoTe_2$ by polarization-resolved second harmonic generation and raman scattering. ACS Nano 2016; **10**: 9626-9636.
10. Janisch C, Wang Y, Ma D, Mehta N, Elias AL, Perealopez N, Terrones M, Crespi VH, Liu Z. Extraordinary second harmonic generation in tungsten disulfide monolayers. Sci. Rep. 2014; **4**: 5530.
11. Zhou X, Cheng J, Zhou Y, Cao T, Hong H, Liao Z, Wu S, Peng H, Liu K, Yu D. Strong second-harmonic generation in atomic layered GaSe. J. Am. Chem. Soc. 2015; **137**: 7994-7997.
12. Yin X, Ye Z, Chenet D, Ye Y, Obrien KJ, Hone J, Zhang X. Edge nonlinear optics on a $MoS_2$ atomic monolayer. Science 2014; **344**: 488-490.
13. Cheng J, Jiang T, Ji Q, Zhang Y, Li Z, Shan Y, Zhang Y, Gong X, Liu W, Wu S. Kinetic nature of grain boundary formation in as-grown $MoS_2$ monolayer. Adv. Mat. 2015; **27**: 4069-4074.
14. Seyler KL, Schaibley JR, Gong P, Rivera P, Jones AM, Wu S, Yan J, Mandrus D, Yao W, Xu X. Electrical

control of second-harmonic generation in a WSe$_2$ monolayer transistor. Nature Nanotech. 2015; **10**: 407-411.
15. Xie C, Mak C, Tao X, Yan F. Photodetectors based on two-dimensional layered materials beyond graphene. Adv. Funct. Mater. 2016; DOI:10.1002/adfm.201603886.
16. Ross JS, Klement P, Jones AM, Ghimire NJ, Yan J, Mandrus D, Taniguchi T, Watanabe K, Kitamura K, Yao W, Cobden DH, Xu X. Electrically tunable excitonic light-emitting diodes based on monolayer WSe$_2$ p-n junctions. Nature Nanotech. 2014; **9**: 268-272.
17. Withers F, Pozozamudio OD, Schwarz S, Dufferwiel S, Walker PM, Godde T, Rooney AP, Gholinia A, Woods CR, Blake P, Haigh SJ, Watanabe K, Taniguchi T, Aleiner IL, Geim AK, Falko VI., Tartakovskii AI, Novoselov KS. WSe$_2$ light-emitting tunneling transistors with enhanced brightness at room temperature. Nano Lett. 2015; **15**: 8223-8228.
18. Gan X, Shiue R, Gao Y, Meric I, Heinz TF, Shepard KL, Hone J, Assefa S, Englund D. Chip-integrated ultrafast graphene photodetector with high responsivity. Nature Photon. 2013; **7**: 883-887.
19. Pospischil A, Humer M, Furchi MM, Bachmann D, Guider R, Fromherz T, Mueller T. CMOS-compatible graphene photodetector covering all optical communication bands. Nature Photon. 2013; **7**: 892-896.
20. Wang X, Cheng Z, Xu K, Tsang HK, Xu J. High-responsivity graphene/silicon-heterostructure waveguide photodetectors. Nature Photon. 2013; **7**: 888-891.
21. Liu M, Yin X, Ulinavila E, Geng B, Zentgraf T, Ju L, Wang F, Zhang X. A graphene-based broadband optical modulator. Nature 2011; **474**: 64-67.
22. Guo J, Xie JJ, Li DJ, Yang GL, Chen F, Wang CR, Zhang LM, Andreev YM, Kokh KA, Lanskii GV, Svetlichnyi VA. Doped GaSe crystals for laser frequency conversion. Light: Science & Applications 2015; **4**: e362.
23. Jiang T, Liu H, Huang D, Zhang S, Li Y, Gong X, Shen Y, Liu W, Wu S. Valley and band structure engineering of folded MoS$_2$ bilayers. Nature Nano. 2014; **9**: 825-829.
24. Miller S, Luke K, Okawachi Y, Cardenas J, Gaeta A, Lipson M. On-chip frequency comb generation at visible wavelength via simultaneous second- and third-order optical nonlinearities. Opt. Express 2014; **22**: 26517-26525.
25. Gan X, Yao X, Shiue R, Hatami F, Englund D. Photonic crystal cavity-assisted upconversion infrared photodetector. Opt. Express 2015; **23**: 12998.
26. Guo X, Zou C, Shuck C, Jung H, Tang H. A parametric down conversion photon-pair source on a silicon chip platform. Light: Science & Applications 2017; **6**: e16249.
27. Castellanos-Gomez A, Buscema M, Molenaar R, Singh V, Janssen L, Zant HSJVD, Steele GA. Deterministic transfer of two-dimensional materials by all-dry viscoelastic stamping. 2D Mater. 2014; **1**: 011002.
28. Tang Y, Mandal KC, McGuire JA, Lai CW. Layer- and frequency-dependent second harmonic generation in reflection from GaSe atomic crystals. Phys. Rev. B 2016; **94**: 125302.
29. Buckley S, Radulaski M, Petykiewicz J, Lagoudakis KG, Kang J, Brongersma ML, Biermann K, Vuckovic J. Second-harmonic generation in GaAs photonic crystal cavities in (111)B and (001) crystal orientations. ACS Photon. 2014; **1**: 516.
30. Rivoire K, Lin Z, Hatami F, Masselink WT, Vuckovic J. Second harmonic generation in gallium phosphide photonic crystal nanocavities with ultralow continuous wave pump power. Opt. Express 2009; **17**: 22609.
31. Galli M, Gerace D, Welna K, Krauss TF, Ofaolain L, Guizzetti G, Andreani LC. Low-power continuous-wave generation of visible harmonics in silicon photonic crystal nanocavities. Opt.

Express 2010; **18**: 26613-26624.
32. Fryett T, Seyler K, Zheng J, Liu C, Xu X, Majumdar A. Silicon photonic crystal cavity enhanced second-harmonic generation from monolayer $WSe_2$. 2D Mater. 2016; **4**: DOI: 10.1088/2053-1583/4/1/015031.
33. Narimatsu M, Kita S, Abe H, Baba T. Array integration of thousands of photonic crystal nanolasers. Appl. Phys. Lett. 2012; **12**: 121117.
34. Gan X, Pervez N, Kymissis I, Hatami F, Englund D. A high-resolution spectrometer based on a compact planar two dimensional photonic crystal cavity array. Appl. Phys. Lett. 2012; **100**: 231104.
35. Gan X, Mak KF, Gao Y, You Y, Hatami F, Hone J, Heinz TF, Englund D. Strong enhancement of light-matter interaction in graphene coupled to a photonic crystal nanocavity. Nano Lett. 2012; **12**: 5626-5631.
36. Yi F, Ren M, Reed JC, Zhu H, Hou J, Naylor CH, Johnson ATC, Agarwal R, Cubukcu E. Optomechanical enhancement of doubly resonant 2D optical nonlinearity. Nano Lett. 2016; 16: 1631-1636.

# Microwatts continuous-wave pumped second harmonic generation in few- and mono-layer GaSe

Xuetao Gan[1,*], Chenyang Zhao,[1] Siqi Hu,[1] Tao Wang,[2] Yu Song,[1] Jie Li,[2] Qinghua Zhao,[2] Wanqi Jie,[2] and Jianlin Zhao[1,*]

[1]*MOE Key Laboratory of Space Applied Physics and Chemistry, and Shaanxi Key Laboratory of Optical Information Technology, School of Science, Northwestern Polytechnical University, Xi'an 710072, China*
[2]*State Key Laboratory of Solidification Processing, Northwestern Polytechnical University, Xi'an 710072, China*
E-mail: xuetaogan@nwpu.edu.cn; jlzhao@nwpu.edu.cn

# Supporting Information

## Fabrication of the integrated GaSe-PC cavity

The PC cavities are fabricated on a silicon-on-insulator (SOI) wafer with a 220 nm thick top silicon layer. The PC cavity patterns are defined on a spin-coated electron-beam resist using the electron beam lithography, which are then transferred into the top silicon layer with an inductively coupled plasma etching. After removing the residual resist layer and wet undercutting the bottom oxide layer using a diluted hydrofluoric acid, air-suspended cavities are obtained.

The lattice structure of the PC cavity is designed with a lattice constant of $a$=450 nm, and air hole radius of $r$=0.25$a$. The defect is formed by shifting the air-holes vertically and horizontally at the cavity center with a distances of 0.2$a$ and 0.12$a$ [1], respectively. To assist the vertical coupling of the cavity mode, the air-holes around the cavity defect are shrunk into a radius of 0.2$a$, as shown in the Fig. 1(c) of the maintext. This cavity design on the 220 nm thick silicon slab could yield resonant wavelengths between 1500 and 1600 nm. Simulated with a finite element method (COMSOL Multiphysics software), the designed cavities could have quality ($Q$) factors as high as 9,000. Unfortunately, due to the fabrication imperfections, the measured $Q$ factors of the fabricated PC cavities are around 2,000.

The bulk GaSe single-crystal is grown by the Bridgeman method. To assist its integration with a PC cavity, few-layer GaSe flakes are mechanically exfoliated onto a polydimethylsiloxane stamp using the Scotch tape, which are then dry transferred onto the cavity with the help of a micromanipulation system [2]. The high quality GaSe crystal enables the easy exfoliation of large few-layer GaSe flakes with dimensions over 1000 μm$^2$ to cover the cavity completely and precisely, as shown in Fig. 1(a) of the maintext.

## Experimental arrangement

The experimental measurements over the GaSe-PC cavity are implemented using a vertically coupled cross-polarization microscope [3], as schematically shown in Fig. S1. The objective lens of the microscope is a near-infrared anti-reflective one with a 50× magnification and a numerical aperture of 0.42. A visible-range white light source and a silicon CCD are employed to monitor the GaSe-PC cavity position during the measurements.

To examine the resonant peak of the PC cavity, the cavity reflection is acquired with an external excitation light source. A polarized beam splitter (PBS) is employed in the microscope system to achieve orthogonally polarized excitation laser and collection signal, allowing for distinguishing the reflection of the cavity mode with a high signal to noise ratio. A half wave plate (HWP) is inserted between the PBS and the objective lens to control the direction of laser polarization with respect to the axis of the GaSe-PC cavity. The employed excitation light source is a narrowband tunable telecom CW laser (Yenista, T100S-HP/CL). The cavity reflection is vertically collected by the objective lens and then reflected by a short-pass dichroic mirror (cutoff wavelength of 1000 nm) into a telecom photodiode. By tuning the laser wavelength with a step of 0.005 nm, the cavity's reflection spectra are obtained, as shown in Fig. 1(d) of the maintext.

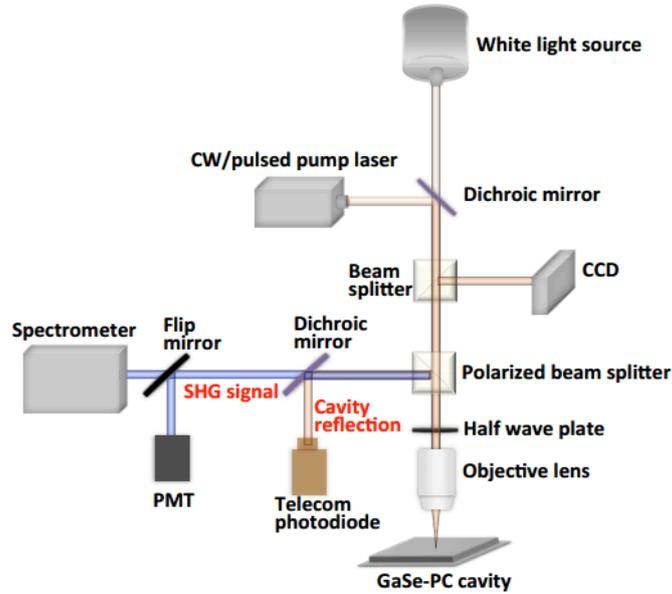

Fig. S1 Schematic diagram of the cross-polarization microscope.

For the measurements of SHG from the GaSe-PC cavity, the CW laser is tuned on-resonance (at the wavelength of 1551 nm) to excite the cavity resonant mode. The frequency conversion signals scattered from the PC cavity are collected by the objective lens, which pass through the short-pass dichroic mirror and are examined by a 0.5 m spectrometer mounted with a cooled silicon camera. As indicated by the Fig. 2(a) of the maintext, SHG and third harmonic generation are obtained. The SHG signal could be redirected into a photomultiplier tube (PMT) as well by a flipping mirror for measuring the SHG power. To evaluate the cavity enhancement factor over the SHG, the CW pump laser is switched into a pulsed picosecond laser (PriTel FFL-20MHz), which is off-resonance from the cavity mode (at the wavelength of 1560 nm).

## Estimation of the SHG enhancement factor by the PC cavity

Due to the optical losses from various optics of the measurement setup as well as the optical absorption by the silicon substrate, it is complicated to measure the absolute SHG power. To evaluate the SHG enhancement factor by the PC cavity, we switch the CW laser into a pulsed laser to pump the GaSe-PC cavity, whose wavelength (at 1560 nm) is off-resonance from the cavity mode. In this regime, both the on-resonance CW pump and the off-resonance pulsed pump share the same excitation and collection optical paths, as well as the same location of GaSe layer to maintain the dielectric environment of the SHG emission. For a 2.8 nW SHG measured by the PMT, the required power for the CW laser and pulsed laser are measured as 2.5 mW and 1.56 mW (averaged power), respectively.

The pulsed laser has Gaussian pulses with a width $\tau$=8.8 ps and a repetition rate $Rep$ = 18.5 MHz. Assuming the Gaussian pulse has a complex envelope with constant phase and Gaussian magnitude, the time-function of the pulse power could be written as

$$P(t) = P_0 \exp(-2t^2/\tau^2) \quad (1)$$

where $P_0$ is the peak power of a single pulse. For the measured averaged power of $P_{ave}$=1.56 mW, the corresponding peak power $P_0$ is 7.6 W, which is calculated by

$$P_0 = \frac{P_{ave} \times 1s}{Rep \times \int P(t) dt} \quad (2)$$

In the second order nonlinear process, the SHG power $P_{SHG}$ is proportional to the squared pump power $P_{pump}$, i.e., $P_{SHG} \propto P^2_{pump}$, and for the pulsed laser should be proportional to $P^2(t)$. For the employed pulsed laser, in a time duration of 1 s, the energy of SHG is proportional to $Rep \times \int P^2(t) dt$. Combining with Eqs. (1)-(2) and $P_{ave}$=1.56 mW, the SHG energy in 1 s generated by the pulsed laser corresponds to a CW pump power of $\sqrt{Rep \times \int P^2(t) dt}/1s = 91.8$ mW.

For the 2.5 mW on-resonance CW pump laser, a power of 0.15 mW couples into the PC cavity with a 6% coupling-in efficiency, which is responsible to the SHG. On the other hand, SHG due to the 1.56 mW off-resonance pulsed laser, corresponds to a CW pump power of 91.8 mW. We therefore calculate the enhancement factor of SHG by the PC cavity as 91.8/0.15=612. Here, we neglect the factors of the focused spot area of the pulsed laser and the in-plane area of the resonant mode, which have the similar values.

## Nonlinear polarizations in GaSe generated by the cavity mode

We define the two-dimensional GaSe lattice (top view) in a $X$-$Y$ coordinate, as schematically shown in Fig. S2(a). Here, the $Y$-axis is along the Ga-Se bond. In a second order nonlinear material, for the electrical fields of a fundamental pump light [$E_X(\omega)$, $E_Y(\omega)$, $E_Z(\omega)$], the induced nonlinear polarizations of SHG [$P_X(2\omega)$, $P_Y(2\omega)$, $P_Z(2\omega)$] could be calculated with the assistance of the second-order susceptibility matrix **d**. The employed $\varepsilon$-GaSe crystal belongs to the $D_{3h}$ symmetry group, and we expect only one independent nonvanishing element of the second-order nonlinear susceptibility ($d_{22}$) [4, 5]. The matrix calculation is described by

$$\begin{bmatrix} P_X(2\omega) \\ P_Y(2\omega) \\ P_Z(2\omega) \end{bmatrix} = \begin{bmatrix} 0 & 0 & 0 & 0 & 0 & -d_{22} \\ -d_{22} & d_{22} & 0 & 0 & 0 & 0 \\ 0 & 0 & 0 & 0 & 0 & 0 \end{bmatrix} \begin{bmatrix} E_X(\omega)E_X(\omega) \\ E_Y(\omega)E_Y(\omega) \\ E_Z(\omega)E_Z(\omega) \\ 2E_Y(\omega)E_Z(\omega) \\ 2E_X(\omega)E_Z(\omega) \\ 2E_X(\omega)E_Y(\omega) \end{bmatrix}.$$

The in-plane PC lattice is defined in a $x$-$y$ coordinate, as displayed in Fig. S2(b). The $x$-$y$ and $X$-$Y$ coordinates are assumed to have an angle of $\phi$ (shown in Fig. S2(c)). On the GaSe-PC cavity, from the resonant mode simulation, we can obtain $E_x(\omega)$ and $E_y(\omega)$ of the resonant mode at the GaSe layer, as shown in Fig. S2(d). Determined by the transverse electrical cavity mode, the vertical component equals to zero, i.e., $E_z(\omega)=0$. The electrical fields of the cavity mode could be transformed into the components in the $X$-$Y$ coordinates, represented as

$$\begin{bmatrix} E_X(\omega) = E_x(\omega)\cos\phi + E_y(\omega)\sin\phi \\ E_Y(\omega) = -E_x(\omega)\sin\phi + E_y(\omega)\cos\phi \\ E_Z(\omega) = 0 \end{bmatrix}.$$

Substituting them into the above matrix calculation, we obtain

$$\begin{bmatrix} P_X(2\omega) = d_{22}\left[(E_x(\omega)^2 - E_y(\omega)^2)\sin(2\phi) + 2E_x(\omega)E_y(\omega)\cos(2\phi)\right] \\ P_Y(2\omega) = d_{22}\left[(E_y(\omega)^2 - E_x(\omega)^2)\cos(2\phi) - 2E_x(\omega)E_y(\omega)\sin(2\phi)\right] \\ P_Z(2\omega) = 0 \end{bmatrix},$$

which are responsible to the measured SHG along the $X$- and $Y$- directions, respectively.

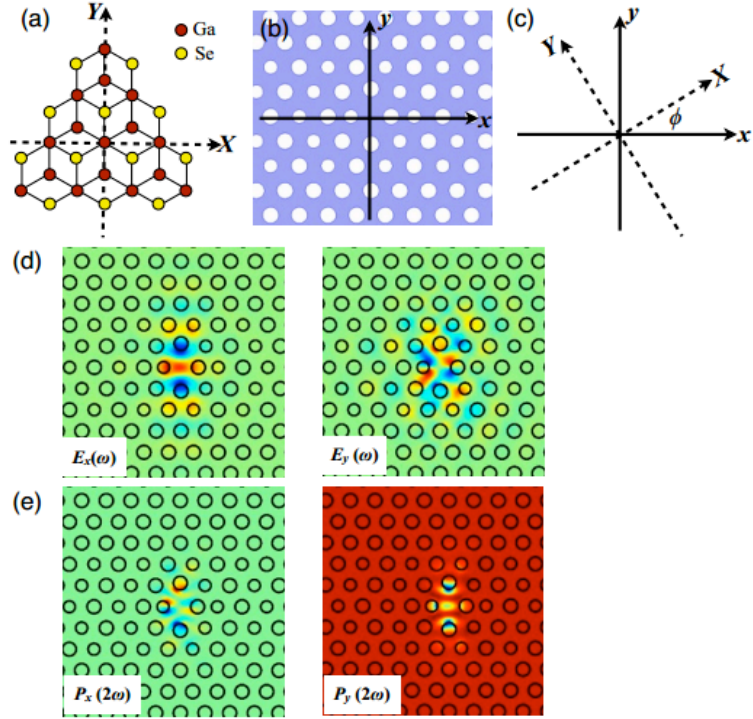

Fig. S2 (a) Two-dimensional GaSe lattice (top view) defined in a $X$-$Y$ coordinate, where the $Y$-axis is along the Ga-Se bond. (b) In-plane PC lattice defined in a $x$-$y$ coordinate. (c) An angle $\phi$ between the $x$-$y$ and $X$-$Y$ coordinates. (d) Simulated electric fields ($E_x(\omega)$ and $E_y(\omega)$) of the resonant mode. (e) Nonlinear polarizations ($P_x(2\omega)$ and $P_y(2\omega)$) calculated from the nonlinear susceptibility matrix and crystal orientation of GaSe.

In our cross-polarization microscope setup, the polarizations of the pump laser excitation and SHG signal collection are selected by the HWP and PBS. Because the crystal orientation of the transferred GaSe flake is unknown, we implement the measurements with respect to the physical orientation of the PC cavity, i.e., $x$-$y$ coordinate. Therefore, we would like transform $P_X(2\omega)$ and $P_Y(2\omega)$ into $P_x(2\omega)$ and $P_y(2\omega)$, respectively, to match our measurement, presenting results of

$$\begin{bmatrix} P_x(2\omega) = (E_x(\omega)^2 - E_y(\omega)^2)\sin\phi + 2E_x(\omega)E_y(\omega)\cos\phi \\ P_y(2\omega) = (E_x(\omega)^2 - E_y(\omega)^2)\cos\phi - 2E_x(\omega)E_y(\omega)\sin\phi \end{bmatrix}.$$

As demonstrated in our maintext, the HWP angle-dependences of the GaSe SHG pumped by the on-resonance CW laser and off-resonance pulsed laser indicate an angle of $\phi=6°$. We can therefore calculate the generated nonlinear polarizations $P_x(2\omega)$ and $P_y(2\omega)$ by the cavity modes $E_x(\omega)$ and $E_y(\omega)$, as shown in Fig. S2(e).

# References

1. Narimatsu M, Kita S, Abe H, Baba T. Enhancement of vertical emission in photonic crystal nanolasers. Appl. Phys. Lett. 2012; 100: 121117.
2. Castellanos-Gomez A, Buscema M, Molenaar R, Singh V, Janssen L, Zant HSJVD, Steele GA. Deterministic transfer of two-dimensional materials by all-dry viscoelastic stamping. 2D Mater. 2014; 1: 011002.
3. Gan X, Mak KF, Gao Y, You Y, Hatami F, Hone J, Heinz TF, Englund D. Strong enhancement of light-matter interaction in graphene coupled to a photonic crystal nanocavity. Nano Lett. 2012; 12: 5626-5631.
4. Shen YR. The principles of nonlinear optics. Wiley-Interscience: New York, 2003.
5. Boyd RW. Nonlinear Optics, 3rd ed.; Academy Press: San Diego, CA, 2008.